\documentclass[10pt]{article}

\usepackage{amstext,amsmath,amssymb,amsfonts,bbm,slashed}
\usepackage[latin1]{inputenc}
\usepackage{epsfig}
\usepackage{hyperref}
\usepackage{amsthm}
\usepackage{subfigure}
\usepackage{color}
\usepackage{multirow}

\newtheorem{definition}{Definition}

\newtheorem{proposition}{Proposition}

\newcommand{\bea}{\begin{eqnarray}}
\newcommand{\eea}{\end{eqnarray}}
\newcommand{\beq}{\begin{equation}}
\newcommand{\eeq}{\end{equation}}

\newcommand{\bpro}{\begin{pro}}
	\newcommand{\epro}{\end{pro}}
\newcommand{\blem}{\begin{lem}}
	\newcommand{\elem}{\end{lem}}
\newcommand{\bdfn}{\begin{dfn}}
	\newcommand{\edfn}{\end{dfn}}
\newcommand{\bcor}{\begin{cor}}
	\newcommand{\ecor}{\end{cor}}
\newcommand{\bthm}{\begin{thm}}
	\newcommand{\ethm}{\end{thm}}
\newcommand{\bex}{\begin{ex}}
	\newcommand{\eex}{\end{ex}}
\newcommand{\brmk}{\begin{rmk}}
	\newcommand{\ermk}{\end{rmk}}
\newcommand{\bpr}{\begin{pr}}
	\newcommand{\epr}{\end{pr}}

\makeatletter

\begin{document}

	\begin{center}
		
		{\LARGE\bf   Kepler dynamics on a conformable Poisson manifold }

		\vspace{15pt}
		
		{\large   Mahouton Norbert Hounkonnou $^{\dagger} $ and Mahougnon Justin Landalidji $^{	\ddagger} $
		}
		
		\vspace{15pt}
		
		{  \,\,
			International Chair of Mathematical Physics
			and Applications  (ICMPA-UNESCO Chair)\\
			University of Abomey-Calavi, 072 B.P. 50 Cotonou, Republic of Benin\\
		$^{\dagger} $	E-mails: { norbert.hounkonnou@cipma.uac.bj with copy to
				hounkonnou@yahoo.fr} \\
				$\ddagger$ E-mails: { landalidjijustin@yahoo.fr}  
		}

	\end{center}
	\vspace{10pt}
	\begin{abstract}
The problem of Kepler dynamics on a conformable Poisson manifold is addressed The Hamiltonian function is defined and the related Hamiltonian vector field governing the dynamics is derived,  which leads to a modified Newton second law.  Conformable momentum and Laplace-Runge-Lenz vectors are considered,  generating $SO(3),  SO(4),$ and $SO(1, 3)$ dynamical symmetry groups. The corresponding first
Casimir operators of $SO(4)$ and $SO(1, 3)$ are,  respectively, obtained. The recursion operators are constructed and used to compute the integrals of motion in action-angle coordinates.  Main relevant properties
are deducted and discussed.
		
		
	\end{abstract}
\section{Introduction}\label{sec1}

Kepler (1609, 1619) derived his three laws of planetary motion using an empirical approach \cite{mu}. From observations, including those made by Tycho Brahe, Kepler deduced the following consequences: $(i)$ the planets move in ellipses with the Sun at one focus; $(ii)$ the radius vector from the Sun to a planet sweeps out equal areas in equal times; $(iii)$ the square of the orbital period of a planet is proportional to the cube of its semi-major axis.		
In the seventeenth century,  Newton  (1687) proved that a simple inverse square law of force gives rise to all motions in the solar system. 
There is good evidence that Robert Hooke, a contemporary and rival of Newton, had proposed the inverse square law of force before Newton \cite{wes},  but Newton's great achievement was to show that Kepler's laws of motion are a natural consequence of this force,  and that the resulting motion is described by a conic section. 
Moreover, although the Kepler problem was a central theme of analytical dynamics for centuries, addressed by many authors, it continues to be so in the contemporary studies as well, revealing interesting mathematical symmetries.

In the 1900s,  special attention was paid to the study of the quantum Kepler system,  $ e. g.,$
the quantum mechanical system of the hydrogen atom, where hidden symmetry was first discovered (see,  for instance, \cite{len,pa,foc,bar,ban1,ban2,hul} and the references therein).
Since the 1970s, modern differential geometry was applied to Kepler problem ( see for instance, \cite{fra,mos} and the references therein). 
These studies led to the re-investigation of the classical Kepler problem (see \cite{fra,bac,gior,gui} and the references therein), giving rise to  $SO(4)$,  $SO(4,2),$ $SO(3,1),$ $O(4),$ $O(4,1),$ and
$O(4,2),$  symmetry groups.
The classical Kepler problem has merited much investigation because of its role in quantum mechanics as well as its inherent importance in classical mechanics \cite{pri}. 
Three constants of motion arise when solving this problem, such as, the energy, the angular momentum and the Runge-Lenz vector, showing that the Kepler problem is a  completely integrable Hamiltonian system in  Liouville and Poincar\'{e} sense \cite{liou,pau}.

A relevant progress in the analysis of the integrability was the important remark that many of these systems exhibit  Hamiltonian dynamics with respect to two compatible symplectic structures \cite{mag1,gel,vil1,fil1,mar,vil2},  giving rise to a geometrical interpretation of the  recursion operator \cite{lax,hkn1}.  From the Magri works,  it is known that the eigenvalues of the recursion operator of bi-Hamiltonian
systems form a set of pairwise Poisson-commuting invariants \cite{bro4}.
It is,  however,  worth noticing that two kinds of difficulties often occur while investigating these systems: (i) firstly,  it is in general very difficult to
give locally an explicit second Hamiltonian structure for a given integrable Hamiltonian system,  even if it is theoretically always possible in the neighborhood of a regular point of the Hamiltonian \cite{bro2}; (ii) secondly,  the global or semi-local existence of such structures implies very strong conditions,  which are rarely satisfied \cite{fern,bro}.
In 1996,  Brouzet {\it et al.} \cite{bro} defined a weaker notion under the name of quasi-bi-Hamiltonian system 
which relaxes these two difficulties for two degrees of freedom.
In 2016,  Cari\~{n}ena {\it et al.} \cite{car}
investigated some properties of the Kepler problem related to the
existence of quasi-bi-Hamiltonian structures.

With regard to our interest  in conformable properties of  manifolds,  let us remind that  already in 1695, Gottfried Leibniz asked Guillaume l'H\^{o}pital if the (integer) order of derivatives and integrals could be extended \cite{val}.  Was it possible if the order was some irrational, fractional or complex number? 
This idea motivated many mathematicians, physicists and engineers to develop the concept of fractional calculus in diverse fields of science and engineering (see $e.g.,$ \cite{agra,metz,herr}). 
Over four centuries,  many famous mathematicians contributed to the theoretical development of fractional calculus.
It is still nowadays one of the most intensively developing areas of mathematical analysis,  highlighting  several definitions of fractional operators like Riemann-Liouville,  Caputo, Gr\"{u}nwald-Letnikov, Riesz and Weyl, etc., \cite{mac}. 
In particular,  in 2014,  Khalil {\it et al.} \cite{kha} introduced the new fractional derivative,  called conformable fractional derivative,  and integral obeying the Leibniz r and chain rules.  Based on this new fractional derivative,  many investigations have been done  (see $e. g.,$ \cite{chun,chun2,chun3,kha2}).
Specifically,  in 2019,  Chun {\it et al.} \cite{chun3} discussed the fractional classical mechanics,  and applied it to the anomalous diffusion relation from the $\alpha$-deformed Langevin equation.  In the same year,  Khalil {\it et al.} gave the geometric meaning of conformable derivative via fractional cords  \cite{kha2}.

Besides,  in 2019 \cite{hkn2},  we studied the Hamiltonian dynamics of the Kepler problem in a deformed phase space,  by considering the equatorial orbit.  In that work,  we constructed recursion operators,  used it to compute the integrals of motion,  and elucidated the existence of quasi-bi-Hamiltonian structures. 
Two years later,  $i.e.,$ in 2021 \cite{hkn3},  we  constructed a hierarchy of bi-Hamiltonian structures for the Kepler problem in a noncommutative phase space,  and computed conserved quantities using  related master symmetries.
More recently,  in 2022 \cite{hkn5},  we investigated Noether symmetry and recursion operators induced by a conformable Poisson algebra in a Minkowski phase space,  constructed
recursion operators using conformable Schwarzschild and Friedmann-Lemaître-Robertson-Walker (FLRW) metrics,  and discussed their relevant master symmetries.
Our present work addresses the problem of Kepler dynamics on a conformable Poisson manifold (CPM).

The paper is organized as follows.  In Section \ref{sec2},  we
give some basic notions useful for our subsequent development.  In Section
\ref{sec3},  we give the
Hamiltonian function,  symplectic form and vector field describing
the Kepler dynamics  on CPM.  In Section \ref{sec4},  we highlight  the existence of  dynamical symmetry groups $SO(3),$ $SO(4),$ and $SO(1,3)$  in the described setting.  In Section \ref{sec5},  we
construct quasi-Hamiltonian,  quasi-bi-Hamiltonian,  and bi-Hamiltonian systems and  their associated recursion
operators,  and deduce the corresponding constants of motion.  In Section
\ref{sec6},  we end with some concluding remarks.

\section{A quick overview on (bi-)Hamiltonian system  and conformable differential: main definitions and notations} \label{sec2}
A Hamiltonian system is here defined as a triple $(\mathcal{Q}, \omega , H )$,  where $(\mathcal{Q}, \omega)$ is a symplectic manifold and  $H$  is a smooth function generating the energy of the system
on $\mathcal{Q}$,  and called  {\it Hamiltonian} or  {\it Hamiltonian function} \cite{rud}.

Given a general dynamical system defined on the $2n$-dimensional manifold $\mathcal{Q}$ \cite{smir,smir2},  its evolution can be described by the equation
{\small\begin{equation} \label{dyn}
	\dot{x}(t) = X(x), \quad x \in \mathcal{Q}, \quad X \in \mathcal{T}\mathcal{Q}.
	\end{equation}}
If the system \eqref{dyn} admits two different Hamiltonian representations:
{\small\begin{equation*} \label{dyn2}
	\dot{x}(t) = X_{H_{1},H_{2}} = \mathcal{P}_{1}dH_{1} = \mathcal{P}_{2}dH_{2},
	\end{equation*}}
its integrability as well as many other properties are subject to Magri's approach. The bi-Hamiltonian vector field $X_{H_{1},H_{2}}$ is defined by two pairs of Poisson bivectors $\mathcal{P}_{1}, \mathcal{P}_{2}$ and Hamiltonian functions $ H_{1},H_{2}.$ Such a  manifold $\mathcal{Q}$ equipped with two Poisson bivectors is called a double Poisson manifold, and the quadruple $(\mathcal{Q},\mathcal{P}_{1},\mathcal{P}_{2},X_{H_{1},H_{2}} )$ is  called a bi-Hamiltonian system.   $\mathcal{P}_{1}$ and $\mathcal{P}_{2}$ are  two compatible Poisson bivectors with vanishing Schouten-Nijenhuis bracket \cite{du}: 
$$[\mathcal{P}_{1}, \mathcal{P}_{2}]_{NS} = 0.$$

A recursion operator 
$
T : \mathcal{T}\mathcal{Q} \longrightarrow 	\mathcal{T}\mathcal{Q}
$
is defined by 
\begin{align*}
T := \mathcal{P}_{2} \circ \mathcal{P}_{1}^{-1}.
\end{align*}
The scarcity of bi-Hamiltonian systems leads to relax the condition for the vector field being
Hamiltonian to a simpler situation of quasi-Hamiltonian with respect to the second symplectic structure \cite{car}.

A Hamiltonian vector field $Y$ on a symplectic manifold
$(\mathcal{M}, \omega) $ is called quasi-bi-Hamiltonian if there
exist another symplectic structure $\omega_{1}$, and a
nowhere-vanishing function $g$, such that $g Y$ is a Hamiltonian
vector field with respect to $ \omega_{1},$ $i.e.,$
\begin{equation}
\iota_{_{Y}}\omega_{0} = -dH_{0}; \quad
\iota_{_{gY}}\omega_{1} =\iota_{_{Y}}(g\omega_{1})  = -dH_{1},
\end{equation}	
where $H_{0}$ and $H_{1}$ are integrals of motion for the
Hamiltonian vector field $Y.$ $g\omega_{1}$ is not closed in
general.
A consequence of this definition is that the pair
$(\omega_{0}, \omega_{1})$ determines a $(1, 1)$-tensor field $T$
defined as $T := \hat{\omega}_{0}^{-1}\circ \hat{\omega}_{1},$ that
is, $\omega_{0}(Y, X) = \omega_{1}(T Y, X ),$ where $X, \ Y$ are two
Hamiltonian vector fields, and $\hat{\omega} := \iota_{_{Y}}\omega.$

A Noether symmetry is a diffeomorphism 	$\Phi : \mathcal{Q} \longrightarrow  \mathcal{Q} $ such that \cite{rom}:

\begin{equation*}
\  \Phi^{\ast} \omega = \omega,\quad 
\ \Phi^{\ast} H = H.
\end{equation*}
An infinitesimal Noether symmetry is a vector field 	$Y \in \mathfrak{X}( \mathcal{Q})$ (the set of all differentiable vector fields on $ \mathcal{Q}$) such that:
\begin{equation*}
\  \mathcal{L}_{Y} \omega = 0, \quad 
\	 \mathcal{L}_{Y} H = 0. 
\end{equation*}

\begin{definition}
	Consider the map $g$ and its inverse $ g^{-1} $: 
	\begin{align*}
	& g:  \mathbb{R}^{2n}_{\alpha}  \longrightarrow  \mathbb{R}^{2n}  \qquad \qquad \qquad \qquad g^{-1}:  \mathbb{R}^{2n}  \longrightarrow \mathbb{R}^{2n}_{\alpha} \\
	& \ \ \qquad z \longmapsto g(z) = |z|^{\alpha - 1}z = \mathbf{Z} \qquad \qquad \ \mathbf{Z} \longmapsto g^{-1} (\mathbf{Z}) = |\mathbf{Z}|^{(1/\alpha) -1}\mathbf{Z} = z,
	\end{align*}
	where 
	$g(0) = 0,\ g(1) = 1,$ and $ g(\pm \infty) = \pm \infty.$
	Then, for this map, the $\alpha$-addition, $\alpha$-subtraction, $\alpha$-multiplication, and $\alpha$-division are given by:
	\begin{align*}
	& a \oplus_{\alpha} b = |a|a|^{\alpha - 1} + b|b|^{\alpha -1}|^{(1/\alpha) - 1} (a|a|^{\alpha - 1} + b|b|^{\alpha - 1}), \\
	& a \ominus_{\alpha} b = |a|a|^{\alpha - 1} - b|b|^{\alpha -1}|^{(1/\alpha) - 1} (a|a|^{\alpha - 1} - b|b|^{\alpha - 1}),\\
	& a \otimes_{\alpha} b = ab, \\
	& a \oslash_{\alpha} b = \dfrac{a}{b},
	\end{align*}
	where $a,b \in \mathbb{R}^{2n}_{\alpha}$. 
\end{definition}
\begin{definition}
	Let $h$ be a differentiable coordinates function on $\mathbb{R}^{2n}_{\alpha}$.  The conformable differential, also called $\alpha$-differential in the sequel,  with respect to the position $q$ and its associated momentum $p$ is defined by:
	\begin{align*}
	d_{\alpha}: \mathbb{R}^{2n}_{\alpha}  &\longrightarrow  \mathbb{R}^{2n} \cr
	h & \longmapsto d_{\alpha}h:=  \sum_{\mu =1}^{2n}    \alpha |x_{\mu}|^{\alpha -1} \dfrac{\partial}{\partial x_{\mu}}h, \quad ( x_{\nu} = q^{\nu}, x_{\nu + n} = p_{\nu}, \ n = 3, \ \nu = 1,2,3)
	\end{align*}
	satisfying the following properties: 
	\begin{itemize}
		\item[(i)] $d_{\alpha} (ah + bf ) = ad_{\alpha}h + bd_{\alpha}f $ for all $ a,b \in \mathbb{R}$;
		\item[(ii)] $ d_{\alpha}(h^{m}) = m h^{m-1}  d_{\alpha}h,$ for all $m \in \mathbb{R}$;
		\item[(iii)] $ d_{\alpha}(c) = 0,$ for all constant functions $h(q,p) = c$;
		\item[(iv)] $ d_{\alpha}( hf) = hd_{\alpha}f + fd_{\alpha}h $ ;
		\item[(v)] $ d_{\alpha}\bigg(\dfrac{h}{f}\bigg) = \dfrac{fd_{\alpha}h - hd_{\alpha}f  }{f^{2}}$,
		where $f$ and $h$ are two differentiable coordinate functions on $\mathbb{R}^{2n}_{\alpha}$;
		\item [(vi)] $ d^{m}_{\alpha}h = \displaystyle\sum_{\mu_{1}...\mu_{m} =1}^{2n}    \alpha^{m} |x_{\mu_{1}}|^{\alpha -1}...|x_{\mu_{m}}|^{\alpha -1} \dfrac{\partial^{m}h}{\partial x_{\mu_{1}}...\partial x_{\mu_{m}}} dx_{\mu_{1}}...dx_{\mu_{m}}, \quad m \in \mathbb{N}.$ 
	\end{itemize}
\end{definition}

The $\alpha$-differential produces a new deformed phase space called  \textit{$\alpha$-deformed phase space.}
The ordinary differential is obtained for $\alpha = 1$.
Using the $\alpha$-addition and  $\alpha$-subtraction, we obtain the following infinitesimal distance between two points of coordinates $ (x_{i}, ...,x_{n})$ and $( x_{i}\oplus_{\alpha} d_{\alpha}x_{i}, ...., x_{n}\oplus_{\alpha} d_{\alpha}x_{n})$  
\begin{align*}\label{eq8}
d_{\alpha}s = (d^{2}_{\alpha}x_{i} + ....+d^{2}_{\alpha}x_{n})^{\frac{1}{2}}.
\end{align*}

\section{ Conformable Poisson manifold and Kepler  Hamiltonian system }\label{sec3}
Our conformable fractional configuration space  is a manifold $\mathcal{Q} = \mathbb{R}_{\alpha}^{3} \backslash \{0\},$
that is, a three-dimensional real euclidean vector space with the origin removed.   
The cotangent
bundle $\mathcal{T}^{\ast}\mathcal{Q}=\mathcal{Q} \times \mathbb{R}_{\alpha}^{3} $ has a  natural  symplectic
structure, 
$$\omega_{\alpha} :
\mathcal{T}\mathcal{Q} \longrightarrow \mathcal{T}^{\ast}\mathcal{Q},$$ which,  in local coordinates  $(q,p)$, i s given by
\begin{equation}\label{Ksy}
\omega_{\alpha} = \sum_{i = 1}^{3} d_{\alpha}p_{i} \wedge  d_{\alpha}q^{i} = \sum_{\mu = 1}^{3} \alpha^{2}|p_{\mu}|^{\alpha - 1}|q^{\mu}|^{\alpha - 1}dp_{\mu} \wedge  dq^{\mu}.
\end{equation} 
Since $\omega_{\alpha}$ is
non-degenerate, it induces an  inverse map, called a    bivector field $\mathcal{P}_{\alpha}$:
$\mathcal{T}^{\ast}\mathcal{Q} \longrightarrow \mathcal{T}\mathcal{Q}$ (tangent bundle)
defined by
\begin{equation}\label{Kbi}
\mathcal{P}_{\alpha} = \sum_{\mu = 1}^{3} \alpha^{-2}|p_{\mu}|^{1 - \alpha }|q^{\mu}|^{1 -\alpha}\dfrac{\partial}{\partial p_{\mu}} \wedge \dfrac{\partial}{\partial q^{\mu}}
,\quad \omega_{\alpha} \circ \mathcal{P}_{\alpha} = \mathcal{P}_{\alpha} \circ \omega_{\alpha} = 1,
\end{equation}
and used  to construct the  Hamiltonian vector field $ X_{\alpha _{f}}$
of a Hamiltonian function $f$ by the relation
\begin{equation}\label{Kvec}
X_{\alpha _{f}} = \mathcal{P}_{\alpha}df. 
\end{equation} 
Then, the manifold  $\mathcal{Q}$ equipped with $\mathcal{P}_{\alpha}$ is a conformable Poisson manifold (CPM).
Consider now the  Kepler Hamiltonian function $H_{\alpha}$ on  $\mathcal{T}^{\ast}\mathcal{Q}$
\begin{align} \label{ha}
H_{\alpha} = \sum_{i=1}^{3} \dfrac{(\alpha|p_{i}|^{\alpha - 1}p_{i})^{2}}{2m} - \dfrac{k}{r_{\alpha}}, \quad r_{\alpha} = \sqrt{\sum_{i=1}^{3} (\alpha|q^{i}|^{\alpha - 1}q^{i})^{2} }
\end{align} 
in which the dynamical variables
satisfy the  conformable Poisson bracket \cite{hkn5}
\begin{equation} \label{Eq_3_10}
\{f,g\}_{\alpha}:= \sum_{i = 1}^{3}\alpha^{-2}|p_{i}|^{1- \alpha}|q^{i}|^{1 - \alpha } \Bigg(\dfrac{\partial{f}}{\partial{p_{i}}}\dfrac{\partial{g}}{\partial{q^{i}}} - \dfrac{\partial{f}}{\partial{q^{i}}}\dfrac{\partial{g}}{\partial{p_{i}}}\Bigg)
\end{equation}
with respect to the above symplectic form 		$\omega_{\alpha}$.  
Then, we obtain the following Hamilton's equations
\begin{align} 
& \dot{q}^{i} := \{ H_{\alpha}, q^{i} \}_{\alpha} = \frac{\alpha}{m}|p_{i}|^{\alpha - 1}p_{i}|q^{i}|^{1 - \alpha} \\
& \dot{p}_{i} := \{ H_{\alpha}, p_{i}\}_{\alpha}= -\dfrac{\alpha k }{r_{\alpha}^{3}} |q^{i}|^{\alpha - 1}q^{i}|p_{i}|^{ 1 - \alpha },
\end{align}
yielding the conformable Newton second law: 
\begin{equation}
\ddot{q} = - \dfrac{\alpha^{3}k}{mr_{\alpha}^{3}} q^{i} + \dfrac{\alpha^{2}}{m^{2}}(1- \alpha)q^{i}p_{i} \bigg(\dfrac{p_{i}}{q^{i}} \bigg)^{2(\alpha - 1)}. 
\end{equation}
The ordinary Newton second law is obtain for 	$\alpha = 1.$
The $1$-form  $dH_{\alpha} \in \mathcal{T}^{\ast}\mathcal{Q}$ and the  Hamiltonian vector field $X_{H_{\alpha}}$ are given by	
\begin{align} \label{Eq_dh}
dH_{\alpha} &= \alpha^{3}\sum_{\mu=1}^{3} \bigg(\dfrac{1}{m}p_{i}^{2\alpha - 1}dp_{i}  + \dfrac{k }{r_{\alpha}^{3}} (q^{i})^{2\alpha - 1}dq^{i} \bigg) \\
X_{H_{\alpha}} &:= \{ H_{\alpha}, .\}_{\alpha}= \alpha\sum_{\mu=1}^{3} \bigg(\dfrac{p_{i}}{m}|p_{i}|^{\alpha - 1}|q^{i}|^{1-\alpha}\dfrac{\partial}{\partial q^{i}}  - \dfrac{k }{r_{\alpha}^{3}} |q^{i}|^{\alpha - 1}q^{i}|p_{i}|^{ 1 - \alpha }\dfrac{\partial}{\partial p_{i}} \bigg)
\end{align}
satisfying the required condition to be a Hamiltonian system
\begin{equation}
\iota_{_{X_{H_{\alpha}}}} \omega_{\alpha}= -dH_{\alpha}.
\end{equation}
On  $\mathcal{T}^{\ast}\mathcal{Q}$,  $q^{j}$ and 	$p_{i}$ satisfy the following commutation relations:
\begin{align}
&\{ p_{i}, q^{j} \}_{\alpha} =  \alpha^{-2}|p_{i}|^{1- \alpha}|q^{j}|^{1 - \alpha } \delta^{i}_{j}, \quad   \{ q^{j}, p_{i}\}_{\alpha}  =  - \alpha^{-2}|p_{i}|^{1- \alpha}|q^{j}|^{1 - \alpha } \delta^{i}_{j},\\
& \{ p_{i}, p_{j} \}_{\alpha} = 0, \quad   \{ q^{i}, q^{j} \}_{\alpha} = 0, \quad i = 1,2,3. 
\end{align}
We notice that for $\alpha = 1,$ we get the Poisson brackets between  $q^{j}$ and 	$p_{i}$ on the ordinary Poisson manifold ($i.e.,$ the Poisson manifold for $\alpha = 1$). 
Since $\{ p_{i}, p_{j} \}_{\alpha} = \{ q^{i}, q^{j} \}_{\alpha} = 0,$ we can deduce that our $\alpha$-deformation does not change the ordinary algebraic structure ($i.e.,$ the algebraic structure for $\alpha = 1$).

\section{Dynamical symmetry groups on conformable Poisson manifold}\label{sec4}
In our framework, the  angular momentum vector $L^{\alpha}$ and the Laplace-Runge-Lenz (LRL) vector $A^{\alpha}$ are   defined, respectively, by:
\begin{equation}
L^{\alpha} = \alpha^{2} |q|^{\alpha - 1}q\times |p|^{\alpha - 1}p, \quad
A^{\alpha} = \alpha|p|^{\alpha - 1}p \times L_{\alpha}- \dfrac{\alpha mk}{r_{\alpha}}|q|^{\alpha - 1}q,
\end{equation}
with the components below:
\begin{align}
&\begin{cases}
L^{\alpha}_{1} = \alpha^{2} (|q^{2}|^{\alpha - 1}q^{2}|p_{3}|^{\alpha - 1}p_{3} -  |q^{3}|^{\alpha - 1}q^{3}|p_{2}|^{\alpha - 1}p_{2})
\\
L^{\alpha}_{2} = \alpha^{2} (|q^{3}|^{\alpha - 1}q^{3}|p_{1}|^{\alpha - 1}p_{1} -  |q^{1}|^{\alpha - 1}q^{1}|p_{3}|^{\alpha - 1}p_{3})
\\
L^{\alpha}_{3} = \alpha^{2} (|q^{1}|^{\alpha - 1}q^{1}|p_{2}|^{\alpha - 1}p_{2} -  |q^{2}|^{\alpha - 1}q^{2}|p_{3}|^{\alpha - 1}p_{3})
\end{cases};\\
&\begin{cases}
A^{\alpha}_{1} = \alpha(|p_{2}|^{\alpha - 1}p_{2}L^{\alpha}_{3} - |p_{3}|^{\alpha - 1}p_{3}L^{\alpha}_{2}) - \dfrac{\alpha mk}{r_{\alpha}}|q^{1}|^{\alpha - 1}q^{1}
\\
\\
A^{\alpha}_{2} = \alpha(|p_{3}|^{\alpha - 1}p_{3}L^{\alpha}_{1} - |p_{1}|^{\alpha - 1}p_{1}L^{\alpha}_{3}) - \dfrac{\alpha mk}{r_{\alpha}}|q^{2}|^{\alpha - 1}q^{2}
\\
\\
A^{\alpha}_{3} = \alpha(|p_{1}|^{\alpha - 1}p_{1}L^{\alpha}_{2} - |p_{2}|^{\alpha - 1}p_{2}L^{\alpha}_{1}) - \dfrac{\alpha mk}{r_{\alpha}}|q^{3}|^{\alpha - 1}q^{3}.
\end{cases}
\end{align}
The commutation relations 	$\{H_{\alpha}, L^{\alpha}_{i}\}_{\alpha}$ and 	$\{H_{\alpha}, A^{\alpha}_{i}\}_{\alpha}$ give:
\begin{equation}
\begin{cases}
\{H_{\alpha}, L^{\alpha}_{1}\}_{\alpha} = 0
\\
\{H_{\alpha}, L^{\alpha}_{2}\}_{\alpha} = 0
\\
\{H_{\alpha}, L^{\alpha}_{3}\}_{\alpha} = 0

\end{cases}; \quad 
\begin{cases}
\{H_{\alpha}, A^{\alpha}_{1}\}_{\alpha} = 0
\\
\{H_{\alpha}, A^{\alpha}_{2}\}_{\alpha} = 0
\\
\{H_{\alpha}, A^{\alpha}_{3}\}_{\alpha} = 0
\end{cases}
\end{equation}
showing that  the vectors $L^{\alpha}$ and $A^{\alpha}$ are in involution with the Hamiltonian function $H_{\alpha},$  $i.e.,$ 
\begin{equation} \label{cnst}
\{H_{\alpha}, L^{\alpha}\}_{\alpha} = 0, \ \{ H_{\alpha}, A^{\alpha}\}_{\alpha} = 0.
\end{equation}
Therefore,  they are constants of motion or first integrals of  $H_{\alpha}$ on  $\mathcal{T}^{\ast}\mathcal{Q}.$ 
Besides, the functions
$L^{\alpha}_{i}$ and $A^{\alpha}_{i}, i = 1,2, 3,$ are also constant along the orbits of $H_{\alpha},$ which, in $\mathcal{T}^{\ast}\mathcal{Q},$  lie in the inverse image of a value of these functions \cite{lig}.
Thus, the following proposition holds: 
\begin{proposition}
	Let $L^{'\alpha}$ be an integral of motion on $\mathcal{T}^{\ast}\mathcal{Q}.$ Then, under the conformable Poisson bracket, its components  $L^{'\alpha}_{i}$'s generate the Lie algebra $so(3)$ of the group $SO(3)$, $i.e., \  \{L^{'\alpha}_{i} , L^{'\alpha}_{j} \}_{\alpha} =  \varepsilon_{i j h} \alpha^{2} L^{'\alpha}_{h},$ where  $ L_{i}^{'\alpha} = - L_{i}^{\alpha}$ and $\varepsilon_{i j h} \alpha^{2}$ are the structure constants of the Lie algebra, and $\varepsilon_{i j h}$  are Levi-Civita symbols giving by
	$\varepsilon_{i j h} := \dfrac{1}{2}(i-j)(j-h)(h-i),$ $ i, j, h = 1,2,3.$
\end{proposition}

\textbf{Proof.} We have
\begin{align*}
\{L^{\alpha}_{1},L^{\alpha}_{2}\}_{\alpha}&:= \sum_{i = 1}^{3}\alpha^{-2}|p_{i}|^{1- \alpha}|q^{i}|^{1 - \alpha } \Bigg(\dfrac{\partial{L^{\alpha}_{1}}}{\partial{p_{i}}}\dfrac{\partial{L^{\alpha}_{2}}}{\partial{q^{i}}} - \dfrac{\partial{L^{\alpha}_{1}}}{\partial{q^{i}}}\dfrac{\partial{L^{\alpha}_{2}}}{\partial{p_{i}}}\Bigg) \cr
& = - \alpha^{4} (|q^{1}|^{\alpha - 1}q^{1}|p_{2}|^{\alpha - 1}p_{2} -  |q^{2}|^{\alpha - 1}q^{2}|p_{3}|^{\alpha - 1}p_{3})\cr
&=  - \alpha^{2}L^{\alpha}_{3} = - \{L^{\alpha}_{2},L^{\alpha}_{1}\}_{\alpha}.
\end{align*} 
Similarly, we get
\begin{align*}
\{L^{\alpha}_{2},L^{\alpha}_{3}\}_{\alpha}
=   - \alpha^{2}L^{\alpha}_{1} = - \{L^{\alpha}_{3},L^{\alpha}_{2}\}_{\alpha}, \qquad
\{L^{\alpha}_{3},L^{\alpha}_{1}\}_{\alpha} = -\alpha^{2}L^{\alpha}_{2} = - \{L^{\alpha}_{1},L^{\alpha}_{3}\}_{\alpha}. 
\end{align*} 
Putting now $L'^{\alpha}_{i} = - L^{\alpha}_{i},$ we obtain 
\begin{align*}
\{L'^{\alpha}_{1},L'^{\alpha}_{2}\}_{\alpha}
&=   \alpha^{2}L'^{\alpha}_{3} = - \{L'^{\alpha}_{2},L'^{\alpha}_{1}\}_{\alpha}, \quad
\{L'^{\alpha}_{2},L'^{\alpha}_{3}\}_{\alpha} = \alpha^{2}L'^{\alpha}_{1} = - \{L'^{\alpha}_{3},L'^{\alpha}_{2}\}_{\alpha}, \cr 
\{L'^{\alpha}_{3},L'^{\alpha}_{1}\}_{\alpha} &=  \alpha^{2}L'^{\alpha}_{2} = - \{L'^{\alpha}_{1},L'^{\alpha}_{3}\}_{\alpha}.
\end{align*} 
leading to $
\{L'^{\alpha}_{i},L'^{\alpha}_{j}\}_{\alpha}
=  \varepsilon_{i j h} \alpha^{2}L'^{\alpha}_{h},  \quad i,j,h = 1,2,3. $ 
$\hfill{\square}$

Now, let us consider some conformable constant energy hypersurfaces $ \Pi_{\alpha c}$ defined as:    \[\Pi_{\alpha c} := \{(q,p)\in \mathcal{T^{\ast}Q}|
H_{\alpha}(q,p) = \alpha c \},  \alpha c \ \mbox{is a constant} .\] 
Since the $1$-form $dH_{\alpha}$  has no zeroes on $\mathcal{T^{\ast}Q},$ we see that the  conformable constant energy hypersurfaces   $\Pi_{\alpha c}$  are closed submanifolds of $\mathcal{T^{\ast}Q}.$
Moreover,  we define the following open submanifolds of $\mathcal{T^{\ast}Q}:$ 
\[\Pi^{\alpha}_{\tau} :=\bigcup_{\alpha c\gtrless 0} \Pi_{\alpha c}  , \quad  \{\tau = -, +\},\]
where $
\mathcal{T^{\ast}Q} = \Pi^{\alpha}_{-} \cup \Pi^{\alpha}_{0} \cup \Pi^{\alpha}_{+},
\ \Pi^{\alpha}_{-} := \{(q,p)\in \mathcal{T^{\ast}Q}|
H_{\alpha}(q,p) < 0 \},$
$\Pi^{\alpha}_{+} := \{(q,p)\in \mathcal{T^{\ast}Q}|H_{\alpha}(q,p) > 0 \};$     $\Pi^{\alpha}_{0}$  is  the common boundary of
$\Pi^{\alpha}_{-}$  and $\Pi^{\alpha}_{+}.$ 

On  $\Pi^{\alpha}_{\tau},$ we introduce 
\begin{equation}
L_{i}^{'\alpha \tau}:= L^{'\alpha}_{i}|\Pi^{\alpha}_{\tau}, \quad \tau= \pm , \  i = 1,2,3,
\end{equation}
and define a scaled Runge-Lenz-Pauli vector $\hat{\Gamma}^{'\alpha}$ by 

\begin{equation}\label{fix1}
\hat{\Gamma}^{'\alpha}= \begin{cases}
\hat{\Gamma}^{'\alpha-} = - \dfrac{1}{(-2mH_{\alpha})^{1/2}}A_{\alpha}, \ \ if \ \ \tau = -
\\
\hat{\Gamma}^{'\alpha+} = - \dfrac{1}{(2mH_{\alpha})^{1/2}}A_{\alpha}, \ \ if  \ \ \tau = + .
\end{cases}
\end{equation} 
We get $\{ H_{\alpha}, \hat{\Gamma}^{\alpha+}_{i} \}_{\alpha} = \{ H_{\alpha}, \hat{\Gamma}^{'\alpha-}_{i} \}_{\alpha}=0$ and  deduce that $\hat{\Gamma}^{'\alpha +}$ and $\hat{\Gamma}^{'\alpha -}$ are also constants of motion on $\Pi^{\alpha}_{+}$ and $\Pi^{\alpha}_{-},$ respectively. Then, we obtain:
\begin{itemize}
	\item[(i)] A Lie algebra isomorphic to the Lie algebra  $so(4)$ of the Lie group $SO(4)$ for  $ \tau = -:$
	\begin{align}
	\{L^{'\alpha-}_{i} , L^{'\alpha-}_{j} \}_{\alpha} & =  \varepsilon_{i j h} \alpha^{2} L^{'\alpha-}_{h},  \cr
	\{\hat{\Gamma}^{'\alpha -}_{i} , \hat{\Gamma}^{'\alpha -}_{j} \}_{\alpha} &=    \varepsilon_{i j h} \alpha^{2}  L^{'\alpha-}_{h},   \cr
	\{L^{'\alpha-}_{i} ,  \hat{\Gamma}^{'\alpha-}_{j} \}_{\alpha} &= \varepsilon_{i j h}\alpha^{2}\hat{\Gamma}^{'\alpha-} _{h},
	\end{align}
	with the associated generators and the first Casimir operators given,  respectively,  by
	{\small \begin{eqnarray*}
			& & \Phi^{\alpha}_{hj} = \sum_{i = 1}^{3} \varepsilon_{h j i} \alpha^{2}L'^{\alpha-}_{i}, \quad h,j,i = 1,2,3, \\
			& & \Phi^{\alpha}_{h4} = - \Phi^{\alpha}_{4h} =  \alpha^{2}\hat{\Gamma}'^{\alpha-}_{h}, \ \Phi^{\alpha}_{44} = 0, \quad h = 1,2,3, \ \mbox{and} \\
			& & C^{\alpha-}_{1} = \sum_{\nu,\mu = 1}^{4}\Phi^{\alpha}_{\nu \mu}\Phi^{\alpha}_{\nu \mu} = 2\alpha^{4}\sum_{i = 1}^{3}\bigg((L'^{\alpha-}_{i})^{2} + (\hat{\Gamma}^{'\alpha-} _{i})^{2}\bigg).					
	\end{eqnarray*}}	
	\item[(ii)]  A  Lie algebra isomorphic to the Lie algebra $so(1,3)$ of the Lie group $SO(1,3)$ for $ \tau = +:$
	\begin{align}
	\{L^{'\alpha +}_{i} , L^{'\alpha+}_{j} \}_{\alpha} &=  \varepsilon_{i j h} \alpha^{2} L^{'\alpha+}_{h}, \cr 
	\{\hat{\Gamma}^{'\alpha +}_{i} , \hat{\Gamma}^{'\alpha +}_{j} \}_{\alpha} & =   - \varepsilon_{i j h} \alpha^{2}  L^{'\alpha+}_{h},  \cr 
	\{L^{'\alpha+}_{i} , \hat{\Gamma}^{'\alpha+}_{j} \}_{\alpha} & = \varepsilon_{i j h}\alpha^{2}\hat{\Gamma}^{'\alpha+} _{h},
	\end{align}
\end{itemize} 
with the corresponding generators and first Casimir operators provided,  respectively,  by
{\small \begin{eqnarray*}
		& & \Psi^{\alpha}_{hj} = \sum_{i = 1}^{3} \varepsilon_{h j i} \alpha^{2}L'^{\alpha+}_{i}, \quad h,j,i = 1,2,3,  \\
		& & \Psi^{\alpha}_{h4} =  \Psi^{\alpha}_{4h} =  - \alpha^{2}\hat{\Gamma}'^{\alpha+}_{h}, \ \Psi^{\alpha}_{44} = 0, \quad h = 1,2,3, \ \mbox{and}\\ \label{psi}
		& & C^{\alpha+}_{1} = \sum_{\nu,\mu = 1}^{4}\Psi^{\alpha}_{\nu \mu}\Psi^{\alpha}_{\nu \mu} = 2\alpha^{4}\sum_{i = 1}^{3}\bigg((L'^{\alpha+}_{i})^{2} + (\hat{\Gamma}^{'\alpha+} _{i})^{2}\bigg).
\end{eqnarray*}} 

\section{Quasi-Hamiltonian, quasi-bi-Hamiltonian, and bi-Hamiltonian systems}  \label{sec5}
Let us now consider the Hamiltonian function 	$H_{\alpha}$ \eqref{ha} in the spherical-polar coordinates 	$(r, \varphi, \vartheta)$, $r \in (0, \infty), \varphi \in (0, 2\pi)$, $\vartheta \in (0, \pi),$
\begin{align}
H_{\alpha} &= \dfrac{1}{2}mr^{2\alpha} \bigg\{ \bigg(\dfrac{\dot{r}}{r}\bigg)^{2}\bigg((\cos\varphi)^{2\alpha} + (\sin\varphi)^{2\alpha}\bigg) \cr
&+ \dot{\vartheta}^{2}\bigg((\sin\vartheta)^{2(\alpha - 1)}(\cos\vartheta)^{2}[(\cos\varphi)^{2\alpha} + (\sin\varphi)^{2\alpha}] + (\cos\vartheta)^{2(\alpha - 1)}(\sin\vartheta)^{2}\bigg)\cr
& + \dot{\varphi}^{2}(\sin\vartheta)^{2\alpha}\bigg((\sin\varphi)^{2} (\cos\varphi)^{2(\alpha - 1)} + (\cos\varphi)^{2} (\sin\varphi)^{2(\alpha - 1)} \bigg) \cr
&+ \bigg(\dfrac{\dot{r}}{r}\bigg)\dot{\vartheta}\sin(2\vartheta) \bigg((\sin\vartheta)^{2(\alpha - 1)}[(\cos\varphi)^{2\alpha} + (\sin\varphi)^{2\alpha}] - (\cos\vartheta)^{2(\alpha - 1)}   \bigg) \cr
& + \bigg(\dfrac{\dot{r}}{r}\bigg)\dot{\varphi}\sin(2\varphi)(\sin\vartheta)^{2\alpha}\bigg((\sin\varphi)^{2(\alpha - 1)} - (\cos\varphi)^{2(\alpha - 1)}  \bigg) \cr
&+ \dot{\varphi}\dot{\vartheta} (\sin\vartheta)^{2\alpha - 1}\sin(2\varphi) \cos\vartheta \bigg((\sin\varphi)^{2(\alpha - 1)} - (\cos\varphi)^{2(\alpha - 1)}  \bigg)\cr
&- k  r^{-\alpha} \bigg((\sin\vartheta)^{2\alpha}[(\cos\varphi)^{2\alpha} + (\sin\varphi)^{2\alpha}] + (\cos\vartheta)^{2\alpha}\bigg)^{-1/2} 
\bigg\}.
\end{align}
Since the case of a plane orbit is still a valid solution of the Kepler problem \cite{ro1}, in the sequel,  we will consider the special case of equatorial orbits, $i. e., \vartheta = \dfrac{\pi}{2}$  with  $ \alpha \geq 1$, and  introduce  the conformable spherical-polar coordinates $(r_{\alpha},\varphi_{\alpha})$ such that  $ \dfrac{\dot{\varphi}_{\alpha}}{\dot{\varphi}} = \dfrac{(\sin(2\varphi))^{2(\alpha - 1)}}{(\sin(\varphi_{\alpha}))^{4}}, \ \varphi_{\alpha} \in (0,2\pi), \ \mbox{and}  \  r_{\alpha} = r^{\alpha}[(\cos\varphi)^{2\alpha} + (\sin\varphi)^{2\alpha}] \in (0,\infty).$
In this case, the  Hamiltonian function 	$H_{\alpha}$ takes the form
\begin{equation}
H_{\alpha} = \dfrac{1}{2m}p^{2}_{r_{\alpha}} + \dfrac{p^{2}_{\varphi_{\alpha}}}{2mr^{2}_{\alpha}(\sin\varphi_{\alpha})^{4}} - \dfrac{k}{r_{\alpha}},
\end{equation} 
where   $ p^{2}_{r_{\alpha}} = \alpha^{-1}m\dot{r}_{\alpha},   \ \mbox{and}  \  p_{\varphi_{\alpha}} = \dfrac{2^{\alpha - 1}m r^{2}_{\alpha} \dot{\varphi}_{\alpha}(\sin\varphi_{\alpha})^{4}}{\sqrt{(\cos\varphi)^{2\alpha} + (\sin\varphi)^{2\alpha}}}.$

In this new coordinate system $(r_{\alpha},\varphi_{\alpha})$, the symplectic form $\omega_{\alpha}$ and the Poisson bivector $\mathcal{P}_{\alpha}$ are given, respectively, by
\begin{equation}
\omega_{\alpha} = dp_{r_{\alpha}} \wedge dr_{\alpha} + (\sin\varphi_{\alpha})^{-2}dp_{\varphi_{\alpha}} \wedge d\varphi_{\alpha}, \  \mathcal{P}_{\alpha} = \dfrac{\partial}{\partial p_{r_{\alpha}} } \wedge  \dfrac{\partial}{\partial r_{\alpha} }  + (\sin\varphi_{\alpha})^{2} \dfrac{\partial}{\partial p_{\varphi_{\alpha}} } \wedge  \dfrac{\partial}{\partial \varphi_{\alpha} } 
\end{equation}
leading to the  Poisson bracket
\begin{equation}
\{f,g\} = \bigg(\dfrac{\partial f}{\partial p_{r_{\alpha}}}\dfrac{\partial g}{\partial r_{\alpha}} - \dfrac{\partial f}{\partial r_{\alpha}}\dfrac{\partial g}{\partial p_{r_{\alpha}}} \bigg) + (\sin\varphi_{\alpha})^{2}\bigg(\dfrac{\partial f}{\partial p_{\varphi_{\alpha}}}\dfrac{\partial g}{\partial \varphi_{\alpha}} - \dfrac{\partial f}{\partial \varphi_{\alpha}}\dfrac{\partial g}{\partial p_{\varphi_{\alpha}}} \bigg)
\end{equation}
and the Hamiltonian vector field 
\begin{align}
X_{H_{\alpha}} &:= \{H_{\alpha},.\}\cr
& = \dfrac{1}{m}p_{r_{\alpha}}\dfrac{\partial }{\partial r_{\alpha}} + \dfrac{1}{m r^{2}_{\alpha}}\bigg(\dfrac{p^{2}_{\varphi_{\alpha}}}{r_{\alpha}(\sin\varphi_{\alpha})^{4}} - mk\bigg)\dfrac{\partial }{\partial p_{r_{\alpha}}} \cr
& + \dfrac{p_{\varphi_{\alpha}}}{m r_{\alpha}^{2}(\sin\varphi_{\alpha})^{2}}\dfrac{\partial }{\partial \varphi_{\alpha}} + \dfrac{2p^{2}_{\varphi_{\alpha}}\cos\varphi_{\alpha}}{mr^{2}_{\alpha}(\sin\varphi_{\alpha})^{3}}\dfrac{\partial }{\partial p_{\varphi_{\alpha}}}, 
\end{align}
which satisfies the relation $ \iota_{_{X_{H_{\alpha}}}} \omega_{\alpha}= -dH_{\alpha}.$  Hence, 
in the coordinate system $(r_{\alpha},\varphi_{\alpha})$
the triplet   \\ $(\mathcal{T}^{\ast}\mathcal{Q},\omega_{\alpha},H_{\alpha})$ is  a Hamiltonian system.

The quantity $\Theta_{\alpha} = \dfrac{p_{\varphi_{\alpha}}}{(\sin\varphi_{\alpha})^{2}}$ is a natural first integral of $H_{\alpha}$, $i.e.,$ $ \{H_{\alpha},\Theta_{\alpha}\} = 0.$
Then, its associated vector field 
\begin{equation}
X_{\Theta_{\alpha}}:= \{\Theta_{\alpha}, .\} = 	\dfrac{\partial }{\partial \varphi_{\alpha}} + 2 p_{\varphi_{\alpha}}\cot\varphi_{\alpha}\dfrac{\partial }{\partial p_{\varphi_{\alpha}}}
\end{equation}
is a dynamical symmetry for $X_{H_{\alpha}}$, $i.e., [X_{H_{\alpha}}, X_{\Theta_{\alpha}}] = 0.$ 
\begin{proposition}
	The vector field $X_{\Theta_{\alpha}}$ is an infinitesimal Noether symmetry.
\end{proposition} 

\textbf{Proof.} We have
\begin{equation} \label{gs}
\mathcal{L}_{X_{\Theta_{\alpha}}}\omega_{\alpha} = d\iota_{X_{\Theta_{\alpha}}}\omega_{\alpha} +  \iota_{X_{\Theta_{\alpha}}} d\omega_{\alpha} = d\iota_{X_{\Theta_{\alpha}}}\omega_{\alpha} = -d^{2} \Theta_{\alpha} = 0,
\end{equation}
and 
\begin{equation} \label{hs}
\mathcal{L}_{X_{\Theta_{\alpha}}}H_{\alpha}= X_{\Theta_{\alpha}}(H_{\alpha})  = 0
\end{equation}
showing that $X_{\Theta_{\alpha}}$ is  both an infinitesimal geometric symmetry, $i.e.,$ leaving invariant the geometric structure, (the symplectic form $\omega_{\alpha}$ ),  and an infinitesimal Hamiltonian symmetry leaving invariant the dynamics, (the Hamiltonian function $H_{\alpha}$).
Hence, $X_{\Theta_{\alpha}}$ is an infinitesimal Noether symmetry. 
$\hfill{\square}$

\subsection{Quasi-Hamiltonian and quasi-bi-Hamiltonian system}
As in \cite{car}, let us denote by 
\begin{equation}
M_{r_{\alpha_{1}}} = \dfrac{p_{r_{\alpha}}p_{\varphi_{\alpha}}}{m (\sin\varphi_{\alpha})^{2}}, \ M_{r_{\alpha_{2}}} = k - \dfrac{p^{2}_{\varphi_{\alpha}}}{mr_{\alpha}(\sin\varphi_{\alpha})^{4}} , \ N_{\varphi_{\alpha_{1}}} = \cos\varphi_{\alpha}, \  N_{\varphi_{\alpha_{2}}} = \sin\varphi_{\alpha}
\end{equation}
four  real functions which are not in involution with the Hamiltonian function $H_{\alpha}$ $i.e.$,
\begin{align}
& \{H_{\alpha}, M_{r_{\alpha_{1}}}\} = -\gamma_{\alpha}M_{r_{\alpha_{2}}}, \ \{H_{\alpha}, M_{r_{\alpha_{2}}}\} = \gamma_{\alpha}M_{r_{\alpha_{1}}}, \label{e1} \\  & \{H_{\alpha}, N_{\varphi_{\alpha_{1}}} \} = -\gamma_{\alpha}N_{\varphi_{\alpha_{2}}}, \ \{H_{\alpha}, N_{\varphi_{\alpha_{2}}} \} = \gamma_{\alpha}N_{\varphi_{\alpha_{1}}},\label{e2}
\end{align}
where $\gamma_{\alpha} = \dfrac{p_{\varphi_{\alpha}}}{mr^{2}_{\alpha} (\sin\varphi_{\alpha})^{2}}.$
Here, \eqref{e2} represents the behaviour of the angular functions $N_{\varphi_{\alpha_{l}}}, l = 1,2$,  and \eqref{e1} represents  behaviour of the functions $ M_{r_{\alpha_{l}}}$.
Consider now the complexification of the linear space of functions on the conformable Poisson manifold  denoting by $M_{r_{\alpha}} $ and $ N_{\varphi_{\alpha}}$ the complex functions
\begin{align}
M_{r_{\alpha}} = M_{r_{\alpha_{1}}} + i M_{r_{\alpha_{2}}}, \quad 	N_{\varphi_{\alpha}} = N_{\varphi_{\alpha_{1}}} + i N_{\varphi_{\alpha_{2}}}
\end{align}
leading to \begin{align}
\{H_{\alpha}, M_{r_{\alpha}}\} = -i\gamma_{\alpha}M_{r_{\alpha}}, \quad  \{H_{\alpha}, N_{\varphi_{\alpha}} \} = -i\gamma_{\alpha}N_{\varphi_{\alpha}}.
\end{align}
Consequently, the following complex function $B^{\alpha}_{\varsigma\beta} = M_{r_{\alpha}}N^{\ast}_{\varphi_{\alpha}}$  is in involution with the Hamiltonian function $ H_{\alpha},$ $i.e.,$
\begin{align}
\{H_{\alpha},B^{\alpha}_{\varsigma\beta} \} = 0 .
\end{align}
Then, the complex function $B^{\alpha}_{\varsigma\beta}$ is a (complex) constant of the motion  and can be written as $	B^{\alpha}_{\varsigma\beta} = B^{\alpha}_{\varsigma} + i B^{\alpha}_{\beta},$  
where
\begin{align}
& B^{\alpha}_{\varsigma} = Re(B^{\alpha}_{\varsigma\beta}) = k \sin \varphi_{\alpha} - \dfrac{p^{2}_{\varphi_{\alpha}}}{mr_{\alpha} (\sin\varphi_{\alpha})^{3}} + \dfrac{p_{r_{\alpha}}p_{\varphi_{\alpha}}\cos\varphi_{\alpha}}{ m(\sin\varphi_{\alpha})^{2}}, \\
& B^{\alpha}_{\beta} = Im(B^{\alpha}_{\varsigma\beta}) =  k \cos \varphi_{\alpha} - \dfrac{p^{2}_{\varphi_{\alpha}}\cos\varphi_{\alpha}}{mr_{\alpha} (\sin\varphi_{\alpha})^{4}} - \dfrac{p_{r_{\alpha}}p_{\varphi_{\alpha}}}{ m\sin\varphi_{\alpha}}
\end{align}
are	two real first integrals,  $i.e.,$
$
\{H_{\alpha},B^{\alpha}_{\varsigma}\} =  	\{H_{\alpha},B^{\alpha}_{\beta} \} = 0.
$

After computation, we find that the Hamiltonian vector field  $ \tilde{X}_{B^{\alpha}_{\varsigma\beta}} =  \{B^{\alpha}_{\varsigma\beta}, .\}$ of the complex function $B^{\alpha}_{\varsigma\beta}$ takes the form 
\begin{align}
\tilde{	X}_{B^{\alpha}_{\varsigma\beta}} = \tilde{X}_{\alpha_{1}} + \tilde{X}_{\alpha_{2}}, \quad \tilde{X}_{\alpha_{1}} = N^{\ast}_{\varphi_{\alpha}} X_{r_{\alpha}}, \ \tilde{X}_{\alpha_{2}} = M_{r_{\alpha}}X^{\ast}_{\varphi_{\alpha}},
\end{align}
where 
\begin{align}
X_{r_{\alpha}} & := \{M_{r_{\alpha}},.\} \cr
& = \bigg(\dfrac{p_{\varphi_{\alpha}}}{m(\sin\varphi_{\alpha})^{2}}\dfrac{\partial}{\partial r_{\alpha}} + \dfrac{p_{r_{\alpha}}}{m} \dfrac{\partial}{\partial \varphi_{\alpha}} + \dfrac{2p_{r_{\alpha}}p_{\varphi_{\alpha}}\cos\varphi_{\alpha}}{m\sin\varphi_{\alpha}} \dfrac{\partial}{\partial p_{\varphi_{\alpha}}}\bigg) \cr
& - i\bigg(\dfrac{p^{2}_{\varphi_{\alpha}}}{mr^{2}_{\alpha}(\sin\varphi_{\alpha})^{4}}\dfrac{\partial}{\partial p_{r_{\alpha}}} + \dfrac{2p_{\varphi_{\alpha}}}{mr_{\alpha}(\sin\varphi_{\alpha})^{2}} \dfrac{\partial}{\partial \varphi_{\alpha}} + \dfrac{4p^{2}_{\varphi_{\alpha}}\cos\varphi_{\alpha}}{mr_{\alpha}(\sin\varphi_{\alpha})^{3}}\dfrac{\partial}{\partial p_{\varphi_{\alpha}}} \bigg)\\
X^{\ast}_{\varphi_{\alpha}} & := \{N^{\ast}_{\varphi_{\alpha}},.\} \cr
&= (\sin\varphi_{\alpha})^{3}\dfrac{\partial}{\partial \varphi_{\alpha}} + i (\sin\varphi_{\alpha})^{2}\cos\varphi_{\alpha}\dfrac{\partial}{\partial p_{\varphi_{\alpha}}},
\end{align}
affording
\begin{align} \label{eq3}
[X_{H_{\alpha}},\tilde{X}_{\alpha_{1}} ] = -iB^{\alpha}_{\varsigma\beta} X_{\gamma_{\alpha}}, \quad 	[X_{H_{\alpha}},\tilde{X}_{\alpha_{2}} ]= iB^{\alpha}_{\varsigma\beta} X_{\gamma_{\alpha}}, 
\end{align}	
with 
\begin{equation} X_{\gamma_{\alpha}} := \{\gamma_{\alpha},.\} = \dfrac{p_{\varphi_{\alpha}}}{mr^{3}_{\alpha}(\sin\varphi_{\alpha})^{2}}\dfrac{\partial}{\partial p_{r_{\alpha}}} + \dfrac{1}{mr^{2}_{\alpha}(\sin\varphi_{\alpha})^{2}}\dfrac{\partial}{\partial \varphi_{\alpha}} + \dfrac{2p_{\varphi_{\alpha}}\cos\varphi_{\alpha}}{mr^{2}_{\alpha}(\sin\varphi_{\alpha})^{3}}\dfrac{\partial}{\partial p_{\varphi_{\alpha}}}.  \end{equation}
From  \eqref{eq3}, we see that  $\tilde{X}_{\alpha_{1}}$ and $\tilde{X}_{\alpha_{2}}$ are not dynamical symmetries of $X_{H_{\alpha}}.$ 
Then, the vector field $X_{\gamma_{\alpha}}$ represents an obstruction for them to be dynamical
symmetries.
Only when  $\gamma_{\alpha}$ is a numerical constant, $\tilde{X}_{\alpha_{1}}$ and $\tilde{X}_{\alpha_{2}}$ are  dynamical
symmetries of $X_{H_{\alpha}}.$
In the following, let us  introduce the complex $2$-form
$\Omega_{\alpha}$ defined as:
\begin{equation}
\Omega_{\alpha}:= dM_{r_{\alpha}} \wedge dN^{\ast}_{\varphi_{\alpha}}  = \Omega_{\alpha_{1}} + i \Omega_{\alpha_{2}}, 
\end{equation}
where 
\begin{align}
\Omega_{\alpha_{1}} &= -\dfrac{1}{m\sin\varphi_{\alpha}}\bigg(p_{r_{\alpha}} + \dfrac{2p_{\varphi_{\alpha}}\cos\varphi_{\alpha}}{r_{\alpha}(\sin\varphi_{\alpha})^{3}} \bigg)  dp_{\varphi_{\alpha}} \wedge d\varphi_{\alpha} - \dfrac{p_{\varphi_{\alpha}}}{m\sin\varphi_{\alpha}} 	dp_{r_{\alpha}} \wedge d\varphi_{\alpha} \cr
&+ \dfrac{p^{2}_{\varphi_{\alpha}}\cos\varphi_{\alpha}}{mr^{2}_{\alpha}(\sin\varphi_{\alpha})^{4}} dr_{\alpha} \wedge d\varphi_{\alpha}, \\
\Omega_{\alpha_{2}} &=  \dfrac{1}{m(\sin\varphi_{\alpha})^{2}}\bigg(-p_{r_{\alpha}}\cos\varphi_{\alpha} + \dfrac{2p_{\varphi_{\alpha}}}{r_{\alpha}\sin\varphi_{\alpha}} \bigg)  dp_{\varphi_{\alpha}} \wedge d\varphi_{\alpha} - \dfrac{p_{\varphi_{\alpha}}\cos\varphi_{\alpha}}{m(\sin\varphi_{\alpha})^{2}}\cr
& \times 	dp_{r_{\alpha}} \wedge d\varphi_{\alpha}  - \dfrac{p^{2}_{\varphi_{\alpha}}}{mr^{2}_{\alpha}(\sin\varphi_{\alpha})^{3}} dr_{\alpha} \wedge d\varphi_{\alpha}. 
\end{align} 
We notice that the Lie derivatives of the initial symplectic form  $\omega_{\alpha}$ with respect to   $\tilde{X}_{\alpha_{1}}$ and   $\tilde{X}_{\alpha_{2}}$ are given, respectively, by
\begin{equation}
\mathcal{L}_{\tilde{X}_{\alpha_{1}}}\omega_{\alpha} = \Omega_{\alpha} \quad \mbox{and} \quad \mathcal{L}_{\tilde{X}_{\alpha_{2}}}\omega_{\alpha} = - \Omega_{\alpha} 
\end{equation}
leading to 
\begin{equation}
\mathcal{L}_{\tilde{X}_{B^{\alpha}_{\varsigma\beta}}}\omega_{\alpha} = 0.
\end{equation}
Then, the complex vector field  $\tilde{X}_{B^{\alpha}_{\varsigma\beta}}$ is an  infinitesimal Noether symmetry. 
\begin{proposition}
	The Hamiltonian vector field  $X_{H_{\alpha}}$ is a  quasi-Hamiltonian system with respect to the complex $2$-form $\Omega_{\alpha}$. 
\end{proposition}
\textbf{Proof.}
Computing the interior product of the $2$-form $\Omega_{\alpha}$ with respect to the Hamiltonian vector field   $X_{H_{\alpha}},$ we obtain 
\begin{equation}
\iota_{X_{H_{\alpha}}} \Omega_{\alpha} =  -i \gamma_{\alpha}dB^{\alpha}_{\varsigma\beta}
\end{equation}
showing that
the Hamiltonian vector field  $X_{H_{\alpha}}$ is a  quasi-Hamiltonian system with respect to the complex $2$-form $\Omega_{\alpha}$. $\hfill{\square}$
\begin{proposition}
	The Hamiltonian vector field  $X_{H_{\alpha}}$ is a quasi-bi-Hamiltonian system  with respect to the two real 2-forms $(\omega_{\alpha},\Omega_{\alpha_{1}})$ or $(\omega_{\alpha},\Omega_{\alpha_{2}})$.
\end{proposition}
\textbf{Proof.}
The interior products of the $2$-forms $\Omega_{\alpha_{1}}$ and $\Omega_{\alpha_{2}}$ with respect to the Hamiltonian vector field   $X_{H_{\alpha}}$ give
\begin{equation}
\iota_{X_{H_{\alpha}}} \Omega_{\alpha_{1}} =  - \gamma_{\alpha}dB^{\alpha}_{\beta}, \quad \iota_{X_{H_{\alpha}}} \Omega_{\alpha_{2}} =   \gamma_{\alpha}dB^{\alpha}_{\varsigma}
\end{equation}
what means that
the Hamiltonian vector field  $X_{H_{\alpha}}$ is also  quasi-bi-Hamiltonian system  with respect to the two real 2-forms $(\omega_{\alpha},\Omega_{\alpha_{1}})$ or $(\omega_{\alpha},\Omega_{\alpha_{2}})$.   $\hfill{\square}$

Besides, we obtain the following associated $\Omega_{\alpha_{i}}$-weaker  recursion operators
{\small\begin{align}
	T_{\alpha_{1}}&:= \omega_{\alpha}^{-1}\circ 	\Omega_{\alpha_{1}} \cr
	& = \bigg[ \dfrac{\partial}{\partial p_{r_{\alpha}} } \wedge  \dfrac{\partial}{\partial r_{\alpha} }  + (\sin\varphi_{\alpha})^{2} \dfrac{\partial}{\partial p_{\varphi_{\alpha}} } \wedge  \dfrac{\partial}{\partial \varphi_{\alpha} } \bigg] \circ \bigg[ \dfrac{-1}{m\sin\varphi_{\alpha}}\bigg(p_{r_{\alpha}} + \dfrac{2p_{\varphi_{\alpha}}\cos\varphi_{\alpha}}{r_{\alpha}(\sin\varphi_{\alpha})^{3}} \bigg)   \cr 
	& \times dp_{\varphi_{\alpha}} \wedge d\varphi_{\alpha} - \dfrac{p_{\varphi_{\alpha}}}{m\sin\varphi_{\alpha}}	dp_{r_{\alpha}} \wedge d\varphi_{\alpha} 
	+ \dfrac{p^{2}_{\varphi_{\alpha}}\cos\varphi_{\alpha}}{mr^{2}_{\alpha}(\sin\varphi_{\alpha})^{4}} dr_{\alpha} \wedge d\varphi_{\alpha} \bigg] \cr
	& = \dfrac{p^{2}_{\varphi_{\alpha}}\cos\varphi_{\alpha}}{mr^{2}_{\alpha}(\sin\varphi_{\alpha})^{4}} \dfrac{\partial}{\partial p_{r_{\alpha}} }\otimes d\varphi_{\alpha} + \dfrac{p_{\varphi_{\alpha}}}{m\sin\varphi_{\alpha}} \dfrac{\partial}{\partial r_{\alpha} } \otimes dp_{r_{\alpha}}  + \dfrac{p_{\varphi_{\alpha}}\sin\varphi_{\alpha}}{m}\dfrac{\partial}{\partial p_{\varphi_{\alpha}} }\otimes dp_{r_{\alpha}} \cr 
	& + \dfrac{\sin\varphi_{\alpha}}{m}\bigg(p_{r_{\alpha}}+ \dfrac{2p_{\varphi_{\alpha}}\cos\varphi_{\alpha}}{r_{\alpha}(\sin\varphi_{\alpha})^{3}} \bigg)\dfrac{\partial}{\partial p_{\varphi_{\alpha}} }\otimes dp_{\varphi_{\alpha}} - \dfrac{p^{2}_{\varphi_{\alpha}}\cos\varphi_{\alpha}}{mr^{2}_{\alpha}(\sin\varphi_{\alpha})^{2}} \dfrac{\partial}{\partial p_{\varphi_{\alpha}} }\otimes dr_{\alpha} \cr
	& +  \dfrac{\sin\varphi_{\alpha}}{m}\bigg(p_{r_{\alpha}}+ \dfrac{2p_{\varphi_{\alpha}}\cos\varphi_{\alpha}}{r_{\alpha}(\sin\varphi_{\alpha})^{3}} \bigg)  \dfrac{\partial}{\partial \varphi_{\alpha} } \otimes d\varphi_{\alpha},\\
	T_{\alpha_{2}}&:= \omega_{\alpha}^{-1}\circ 	\Omega_{\alpha_{2}} \cr
	& = \bigg[ \dfrac{\partial}{\partial p_{r_{\alpha}} } \wedge  \dfrac{\partial}{\partial r_{\alpha} }  + (\sin\varphi_{\alpha})^{2} \dfrac{\partial}{\partial p_{\varphi_{\alpha}} } \wedge  \dfrac{\partial}{\partial \varphi_{\alpha} } \bigg] \circ \bigg[  - \dfrac{p_{\varphi_{\alpha}}\cos\varphi_{\alpha}}{m(\sin\varphi_{\alpha})^{2}}	dp_{r_{\alpha}} \wedge d\varphi_{\alpha} \cr
	& + \dfrac{1}{m(\sin\varphi_{\alpha})^{2}}\bigg(-p_{r_{\alpha}}\cos\varphi_{\alpha} + \dfrac{2p_{\varphi_{\alpha}}}{r_{\alpha}\sin\varphi_{\alpha}} \bigg)  dp_{\varphi_{\alpha}} \wedge d\varphi_{\alpha}  - \dfrac{p^{2}_{\varphi_{\alpha}}}{mr^{2}_{\alpha}(\sin\varphi_{\alpha})^{3}} dr_{\alpha} \wedge d\varphi_{\alpha} \bigg] \cr
	& =  - \dfrac{p^{2}_{\varphi_{\alpha}}}{mr^{2}_{\alpha}(\sin\varphi_{\alpha})^{3}}\dfrac{\partial}{\partial p_{r_{\alpha}} }\otimes d\varphi_{\alpha} +  \dfrac{p_{\varphi_{\alpha}}\cos\varphi_{\alpha}}{m(\sin\varphi_{\alpha})^{2}}\dfrac{\partial}{\partial r_{\alpha} } \otimes d\varphi_{\alpha} + \dfrac{p_{\varphi_{\alpha}}\cos\varphi_{\alpha}}{m}\dfrac{\partial}{\partial p_{\varphi_{\alpha}} }\otimes dp_{r_{\alpha}} \cr 
	& - \dfrac{1}{m}\bigg(-p_{r_{\alpha}}\cos\varphi_{\alpha} + \dfrac{2p_{\varphi_{\alpha}}}{r_{\alpha}\sin\varphi_{\alpha}} \bigg)\dfrac{\partial}{\partial p_{\varphi_{\alpha}} }\otimes dp_{\varphi_{\alpha}} +  \dfrac{p^{2}_{\varphi_{\alpha}}}{mr^{2}_{\alpha}\sin\varphi_{\alpha}}\dfrac{\partial}{\partial p_{\varphi_{\alpha}} }\otimes dr_{\alpha} \cr
	&  - \dfrac{1}{m}\bigg(-p_{r_{\alpha}}\cos\varphi_{\alpha} + \dfrac{2p_{\varphi_{\alpha}}}{r_{\alpha}\sin\varphi_{\alpha}} \bigg)  \dfrac{\partial}{\partial \varphi_{\alpha} } \otimes d\varphi_{\alpha}.
	\end{align}}
It is important to notice that each recursion operator $T_{\alpha_{i}}$ is not $X_{H_{\alpha}}-$invariant $i.e.,  \mathcal{L}_{X_{H_{\alpha}}}T_{\alpha_{i}} \neq 0$. Then, they do not satisfy the Nijenhuis torsion condition mentioned in  \cite{fil1}. 
\subsection{Bi-Hamiltonian system}
Since the Hamiltonian function $H_{\alpha}$ does not explicitly depend on the time, then setting $V_{\alpha} = W_{\alpha} - E_{\alpha}t,$ it is possible to find an additive separable solution:
\begin{equation}
W_{\alpha} = W_{r_{\alpha}}(r_{\alpha}) +  W_{\varphi_{\alpha}}(\varphi_{\alpha}). 
\end{equation}  
In this case, the Hamilton-Jacobi equation  is  reduced to:
\begin{equation}
E_{\alpha}= \dfrac{1}{2m}\bigg( \dfrac{\partial{W_{\alpha}}}{\partial{r_{\alpha}}}\bigg)^{2}  + \dfrac{1}{2mr_{\alpha}^{2}(\sin\varphi_{\alpha})^{4}}\bigg( \dfrac{\partial{W_{\alpha}}}{\partial{\varphi_{\alpha}}}\bigg)^{2} - \dfrac{k}{r_{\alpha}},
\end{equation}
leading to the following set of equations:
\begin{align*}
& \begin{cases}
\dfrac{1}{(\sin\varphi_{\alpha})^{4}}\bigg( \dfrac{dW_{\varphi_{\alpha}} (\varphi_{\alpha})}{d\varphi_{\alpha}}\bigg)^{2}= \mathcal{D}^{2} \\
-r_{\alpha}^{2}\bigg(\dfrac{dW_{r_{\alpha}}(r_{\alpha})}{dr_{\alpha}}\bigg)^{2} + 2mr_{\alpha}^{2}E_{\alpha} + 2mr_{\alpha}k = \mathcal{D}^{2}, \quad \mathcal{D}^{2} \ \mbox{is a constant} 
\end{cases} \cr
\Rightarrow &\begin{cases}
\bigg( \dfrac{dW_{\varphi_{\alpha}} (\varphi_{\alpha})}{d\varphi_{\alpha}}\bigg)^{2}=(\sin\varphi_{\alpha})^{4}\mathcal{D}^{2}\\
\bigg(\dfrac{dW_{r_{\alpha}}(r_{\alpha})}{dr_{\alpha}}\bigg)^{2}  =  2mE_{\alpha} +  \dfrac{2mk}{r_{\alpha}} - \dfrac{\mathcal{D}^{2}}{r_{\alpha}^{2}}.
\end{cases}
\end{align*}
In the compact case, characterized by $E_{\alpha} < 0,$
we can introduce the action variables 
\begin{equation}
\begin{cases}
J_{\varphi_{\alpha}}= \dfrac{1}{2\pi}\oint \dfrac{dW_{\varphi_{\alpha}}  (\varphi)}{d\varphi_{\alpha}}d\varphi_{\alpha} = \dfrac{1}{2\pi}\oint (\sin\varphi_{\alpha})^{2}\mathcal{D} d\varphi_{\alpha}\\
J_{r_{\alpha}} =\dfrac{1}{2\pi} \oint \dfrac{dW_{r_{\alpha}}(r_{\alpha})}{dr_{\alpha}}dr_{\alpha} = \dfrac{1}{2\pi} \oint \bigg( 2mE_{\alpha} + \dfrac{2mk}{r_{\alpha}} - \dfrac{\mathcal{D}^{2}}{r_{\alpha}^{2}} \bigg)^{1/2}dr_{\alpha}.
\end{cases}
\end{equation}  
and angle variables
\begin{equation} \label{angle}
\phi_{\alpha}^{i} = \dfrac{\partial W_{\alpha}}{\partial J_{\alpha_{i}}}, \quad \phi_{\alpha}^{i}(0) = 0, \quad  i = 1, 2, 3.
\end{equation}
Using the standard integration method, we get
\begin{equation} \label{act-ang}
\begin{cases}
J_{\alpha_{1}} = J_{r_{\alpha}} = - \mathcal{D} + \dfrac{mk}{\sqrt{- 2mE_{\alpha}}}\\
J_{\alpha_{2}} =  J_{\varphi_{\alpha}} = \dfrac{\mathcal{D} }{2} \\
\phi_{\alpha}^{1} = -\dfrac{1}{(J_{\alpha_{1}} + 2J_{\alpha_{2}})} \sqrt{\tilde{G}_{\alpha}} + \arcsin\bigg[ \dfrac{mkr_{\alpha} - (J_{\alpha_{1}} +  2J_{\alpha_{2}})^{2}}{\tilde{Q}_{\alpha}}\bigg]\\
\phi_{\alpha}^{2} =  2\phi_{\alpha}^{1} - 2\arcsin\bigg[\dfrac{ \bigg(1 - \dfrac{(2J_{\alpha_{2}})^{2}}{mkr_{\alpha}}\bigg)(J_{\alpha_{1}} + 2J_{\alpha_{2}})}{\sqrt{(J_{\alpha_{1}} + 2J_{\alpha_{2}})^{2} - (2J_{\alpha_{2}})^{2}}}  \bigg]  + \varphi_{\alpha}  - \dfrac{1}{4}\sin(2\varphi_{\alpha}),
\end{cases}  
\end{equation}
where  \begin{align*}
&\tilde{Q}_{\alpha} = (J_{\alpha_{1}} + 2J_{\alpha_{2}})\sqrt{(J_{\alpha_{1}} + 2J_{\alpha_{2}})^{2} - (2J_{\alpha_{2}})^{2}}, \cr
&\tilde{G}_{\alpha} = -m^{2}k^{2}r_{\alpha}^{2} + 2mk(J_{\alpha_{1}} + 2J_{\alpha_{2}})^{2} r_{\alpha} - (2J_{\alpha_{2}})^{2}(J_{\alpha_{1}} + 2J_{\alpha_{2}})^{2} .  
\end{align*}
Then, we obtain in action-angle coordinates  $(J_{\alpha},\phi_{\alpha}),$
the Hamiltonian  $H_{\alpha},$ the  Poisson bivector  $\mathcal{P}_{\alpha},$ the
symplectic form  $ \omega_{\alpha},$  and the Hamiltonian vector field  $X_{H_{\alpha}},$
\begin{align} \label{Bil0}  
& H_{\alpha} = E_{\alpha} =  -\dfrac{mk^{2}}{2(J_{\alpha_{1}} + 2J_{\alpha_{2}})^{2}},\ \  \mathcal{P}_{\alpha}  = \sum_{h = 1}^{2}\dfrac{\partial}{\partial J_{\alpha_{h}}} \wedge \dfrac{\partial}{\partial\phi_{\alpha}^{h}}, \\
&\omega_{\alpha}  =   \sum_{h = 1}^{2} dJ_{\alpha_{h}}  \wedge d\phi_{\alpha}^{h},\ \
X_{H_{\alpha}} := \{H_{\alpha},.\} =
\dfrac{mk^{2}}{(J_{\alpha_{1}} + 2J_{\alpha_{2}})^{3}}\bigg(
\dfrac{\partial}{\partial{\phi_{\alpha}^{1}}} +
2\dfrac{\partial}{\partial{\phi_{\alpha}^{2}}}  \bigg)
\end{align} 
satisfying $  \iota_{_{X_{H_{\alpha}}}} \omega_{\alpha}= -dH_{\alpha}.$
Therefore, in the action-angle coordinates  $(J_{\alpha},\phi_{\alpha}),$ 
the triplet   $(\mathcal{T}^{\ast}\mathcal{Q},\omega_{\alpha},H_{\alpha})$ is also a Hamiltonian system.

In addition,  let us introduce
the vector field 
\begin{equation}
\Delta^{\alpha} := \displaystyle\sum_{h=1}^{2} \lambda^{\alpha}_{h} \dfrac{\partial}{\partial{J_{\alpha _{h}}}},  
\quad  \left\{
\begin{array}{ll}
\lambda^{\alpha}_{1} = \dfrac{1}{2}\bigg( J_{\alpha _{1}}^{2} + 4J_{\alpha _{2}}^{2} \bigg) \\ 
\lambda^{\alpha}_{2} = J_{\alpha _{1}}J_{\alpha _{2}}.
\end{array} 
\right.
\end{equation}
Since $\Delta^{\alpha}$ is an non-infinitesimal geometric symmetry, $i.e.,$ $\mathcal{L}_{\varDelta^{\alpha}}\omega_{\alpha} = \omega_{\alpha_{1}} \neq 0$, and an non-infinitesimal Hamiltonian symmetry, $i.e.,$ $\varDelta^{\alpha}(H_{\alpha}) = \tilde{H}_{\alpha}\neq 0,$ then it is an non-infinitesimal Noether symmetry. 
Notice that 
\begin{equation}
\omega_{\alpha_{1}} = \sum_{h,k = 1}^{2} (R_{\alpha})_{h}^{k}J_{\alpha _{k}}\wedge d\phi_{\alpha}^{h}, \quad R_{\alpha} =  \left(
\begin{array}{ll}
J_{\alpha _{1}}& J_{\alpha _{2}}   \\ 
4 J_{\alpha _{2}} & J_{\alpha _{1}} 
\end{array}
\right) ,
\end{equation}
is also a symplectic form and the function
\begin{equation}
\tilde{H}_{\alpha} = \dfrac{mk^{2}}{2(J_{\alpha_{1}} + 2J_{\alpha_{2}})}
\end{equation}
is a first integral of the Hamiltonian function $ H_{\alpha}$, $i.e., \{H_{\alpha},\tilde{H}_{\alpha}\} = 0.$	

Now,  taking the vector field $\Delta^{\alpha}$  in the form
$\Delta_{j}^{\alpha}= \dfrac{2}{(3 - j + 1)} \displaystyle\sum_{h=1}^{2} \lambda^{\alpha}_{h} \dfrac{\partial}{\partial{J_{\alpha _{h}}}}, \ j = 1,2,3,$ it
generates a sequence of finitely many  dynamical symmetries according to the following scheme:
\begin{equation}
X_{\alpha_{i + 1}} := [X_{\alpha_{i}}, \Delta_{j}^{\alpha}] = \dfrac{(3 - i)!mk^{2}}{(3 - j)! ( 3 -i)(J_{\alpha _{1}} + 2J_{\alpha _{2}} )^{3-j}}\bigg( \dfrac{\partial}{\partial{\phi_{\alpha}^{1}}} + 2\dfrac{\partial}{\partial{\phi_{\alpha}^{2}}}\bigg), 
\end{equation}
where $X_{\alpha_{i}} = 	\dfrac{mk^{2}}{(J_{\alpha_{1}} + 2J_{\alpha_{2}})^{3 -i}}\bigg(
\dfrac{\partial}{\partial{\phi_{\alpha}^{1}}} +
2\dfrac{\partial}{\partial{\phi_{\alpha}^{2}}}  \bigg),   X_{\alpha_{0}} = X_{H_{\alpha}} , \mbox{and} \ i = j - 1.$
Explicitly, we have:           
\begin{align}
& X_{\alpha_{0}} = \dfrac{mk^{2}}{(J_{\alpha _{1}} + 2J_{\alpha _{2}} )^{3}}\bigg( \dfrac{\partial}{\partial{\phi_{\alpha}^{1}}} + 2 \dfrac{\partial}{\partial{\phi_{\alpha}^{2}}}\bigg), \ X_{\alpha_{1}} = \dfrac{mk^{2}}{(J_{\alpha _{1}} + 2J_{\alpha _{2}})^{2}}\bigg( \dfrac{\partial}{\partial{\phi_{\alpha}^{1}}} + 2 \dfrac{\partial}{\partial{\phi_{\alpha}^{2}}}\bigg) \\
& X_{\alpha_{2}} = \dfrac{mk^{2}}{(J_{\alpha _{1}} + 2J_{\alpha _{2}})}\bigg( \dfrac{\partial}{\partial{\phi_{\alpha}^{1}}} + 2 \dfrac{\partial}{\partial{\phi_{\alpha}^{2}}}\bigg), \quad X_{\alpha_{3}} = mk^{2}\bigg( \dfrac{\partial}{\partial{\phi_{\alpha}^{1}}} + 2 \dfrac{\partial}{\partial{\phi_{\alpha}^{2}}}\bigg),
\end{align}	
which are in involution with each other, $i.e.,$ 
$[X_{\alpha_{h}}, X_{\alpha_{k}}] = 0, \quad h,k = 0,1,2,3.	
$
Hence, each $\Delta_{j}^{\alpha}$ is a master symmetry \cite{hkn3,hkn4,hkn5} for $ X_{H_{\alpha}}$. 
The vector fields $X_{\alpha_{i}}$ are Hamiltonian vector fields, $i. e.,$ can be expressed as: 
\begin{equation} 
X_{\alpha_{i}} = \{H_{\alpha_{i}}, .\} = \{H_{\alpha_{i + 1}}, .\}_{_{1}}, \ H_{\alpha_{0}} = H_{\alpha}, \ i= 0, 1, 2, 
\end{equation}
with respect to  the  Poisson bracket $  \{ .,.\}_{_{1}}$ defined as follows: 					
\begin{equation}
\{ f,g\}_{_{1}} := \sum_{h,k = 1}^{2}(R_{\alpha}^{-1})^{h}_{k} \bigg(\dfrac{\partial{f}}{\partial{J_{\alpha_{k}}}}\dfrac{\partial{g}}{\partial{\phi_{\alpha}^{h}}} - \dfrac{\partial{f}}{\partial{\phi_{\alpha}^{h}}}\dfrac{\partial{g}}{\partial{J_{\alpha_{k}}}}\bigg),
\end{equation}

with: 
$           R_{\alpha}^{-1} = \left ( \begin{array}{ll}
\dfrac{J_{\alpha_{1}}}{(J_{\alpha_{1}} - 2J_{\alpha_{2}}) (J_{\alpha_{1}} + 2J_{\alpha_{2}} )}& \dfrac{- 4J_{\alpha_{2}} }{(J_{\alpha_{1}} - 2J_{\alpha_{2}}) (J_{\alpha_{1}} + 2J_{\alpha_{2}})} \\
\\
\dfrac{- J_{\alpha_{2}} }{(J_{\alpha_{1}} - 2J_{\alpha_{2}}) (J_{\alpha_{1}} + 2J_{\alpha_{2}} )} & \dfrac{ J_{\alpha_{1}} }{(J_{1} - 2J_{\alpha_{2}}) (J_{\alpha_{1}} + 2J_{\alpha_{2}})}
\end{array}\right ),
$
and	      
\begin{align}
&H_{\alpha} = \dfrac{- mk^{2}}{2(J_{\alpha_{1}} + 2J_{\alpha_{2}})^{2}}, 	H_{\alpha_{1}} = \dfrac{-mk^{2}}{(J_{\alpha_{1}} + 2J_{\alpha_{2}})}, \\ &H_{\alpha_{2}} =  mk^{2} \ln(J_{\alpha_{1}} + 2J_{\alpha_{2}}), \ H_{\alpha_{3}} = mk^{2}(J_{\alpha_{1}} + 2J_{\alpha_{2}}). 
\end{align}
\begin{proposition}
	For  $i = 0,1,2,$  each $X_{\alpha_{i}}$ is a bi-Hamiltonian vector field and $(\mathcal{T^{\ast}Q},\mathcal{P}_{\alpha_{1}},\mathcal{P}_{\alpha},X_{\alpha_{i}} )$ is a bi-Hamiltonian system, where   \begin{equation}\mathcal{P}_{\alpha_{1}} = \displaystyle\sum_{h,k = 1}^{2} (R^{-1}_{\alpha})_{k}^{h} \dfrac{\partial}{\partial J_{\alpha_{h}}} \wedge \dfrac{\partial}{\partial\phi_{\alpha}^{k}},	\end{equation}   
	{\it i.e.,}
	\begin{equation*}
	\iota_{_{X_{\alpha_{i}}}} \omega_{\alpha} = - dH_{\alpha_{i}} \quad \mbox{and} \quad \iota_{_{X_{\alpha_{i}}}} \omega_{\alpha_{1}} = - dH_{\alpha_{i + 1}}, \quad  X_{\alpha_{i}} = \{H_{\alpha_{i}}, .\} = \{H_{\alpha_{i + 1}}, .\}_{_{1}},
	\end{equation*}
	where $H_{\alpha_{i}}  (H_{\alpha_{0}}\equiv H_{\alpha})$ are integrals of motion for $X_{\alpha_{i}}.$
\end{proposition}
\textbf{Proof.} Since
{\small\begin{equation*}
	\iota_{X}(df \wedge dg) = (Xf)dg - (df)Xg,
	\end{equation*}}
we obtain
{\small\begin{equation*}
	\iota_{_{X_{\alpha_{i}}}} \omega_{\alpha} = - dH_{\alpha_{i}} \quad \mbox{and} \quad \iota_{_{X_{\alpha_{i}}}} \omega_{\alpha_{1}} = - dH_{\alpha_{i + 1}}, \quad  X_{\alpha_{i}} = \{H_{\alpha_{i}}, .\} = \{H_{\alpha_{i + 1}}, .\}_{_{1}}.
	\end{equation*}}
$\hfill{\square}$

Then, the recursion operator for the Kepler dynamics in action-angle coordinates $(J_{\alpha},\phi_{\alpha})$ is given by:
\begin{equation} \label{Ro1}
T_{\alpha} := \mathcal{P}_{\alpha_{1}}\circ \mathcal{P}_{\alpha}^{-1} =  \sum_{h,k =1 }^{2}(R^{-1}_{\alpha})_{k}^{h}\bigg(  \dfrac{\partial}{\partial{J_{\alpha_{h}}}}\otimes dJ_{\alpha_{k}} +  \dfrac{\partial}{\partial{\phi^{h}_{\alpha}}}\otimes d\phi^{k}_{\alpha}\bigg),
\end{equation}
where the components of the matrix $(R^{-1}_{\alpha})$ are constants of motion. Furthermore, we notice that the eigenvalues of $(R^{-1}_{\alpha})$, $I_{1} = J_{\alpha_{1}} -2 J_{\alpha_{1}}$ and  $I_{2} = J_{\alpha_{1}} + 2 J_{\alpha_{1}} $ {are real distinct constants of motion}. Hence, according to Magri \cite{mag1}, these eigenvalues  form a system of first integrals in involution. 
This shows that the Kepler Hamiltonian dynamics hints at a connection between the geometry of our physical system, ($\alpha-$deformed symplectic manifolds and related Hamiltonian vector fields), and conservation laws.
In this connection, the planet positions  on the conformable Poisson manifold are viewed as states and vector fields as laws governing how those states evolve.

\section{Concluding remarks} \label{sec6}
In this paper, we have given the
Hamiltonian function, symplectic form and vector field describing
the Kepler dynamics  on a conformable Poisson manifold,  and proved the existence of dynamical symmetry
groups $SO(3),$ $SO(4),$ and $SO(1,3)$.
In addition, we have investigated the Kepler problem in a conformable spherical-polar coordinates by considering the special case of equatorial orbits, $i. e., \vartheta = \dfrac{\pi}{2}$  with  $ \alpha \geq 1.$ Then, we have obtained a natural constant of motion, highlighted the existence of infinitesimal Noether symmetries,  constructed quasi-Hamiltonian and quasi-bi-Hamiltonian systems, and computed the associated  recursion operators. 

As in \cite{car},  we have got in the considered conformable Poisson manifold that the superintegrability of the
Kepler problem  is directly related with the existence of two complex functions whose conformable Poisson brackets with the Hamiltonian are proportional with a common complex factor to themselves. This investigation is therefore of great importance because it has determined both
the existence of superintegrability (existence of additional constants of motion) and  quasi-bi-Hamiltonian structures. 

Futher, we have studied the same Kepler problem in action-angle  coordinates,  obtained its corresponding Hamiltonian
system,  derived a family of dynamical symmetries $X_{\alpha_{i}}$ leading to the construction of a bi-Hamiltonian structure and its related recursion operator $T_{\alpha},$ which generates constants of motion. This recursion operator is analogue to the recursion operator obtained in \cite{hkn2}. Then,  we can deduce that the tensor $T_{\alpha}$  satisfies  the Nijenhuis torsion condition mentioned in  \cite{fil1}.

\subsection*{Acknowledgments}
The ICMPA-UNESCO Chair is in partnership
with the Association pour la Promotion Scientifique de l'Afrique
(APSA), France, and Daniel Iagolnitzer Foundation (DIF), France,
supporting the development of mathematical physics in Africa.

\end{document}